\documentclass[hidelinks]{article}
\usepackage{cite}

\usepackage[aboveskip=1pt,belowskip=0pt]{subcaption}
\usepackage{spconf,amsmath}
\usepackage{amssymb}
\usepackage{float}
\usepackage{url}
\usepackage[capitalise]{cleveref}
\usepackage{tikz}
\usepackage{circuitikz}
\usetikzlibrary{shapes,arrows,calc,fit,matrix,quotes}
\usepackage{epstopdf}
\newcommand{\emaildot}{\makebox[0.2em]{\scalebox{.23}{\textbullet}}}

\title{Reconstruction of FRI Signals using Deep Neural Network Approaches}
\name{Vincent C. H. Leung\thanks{Vincent C. H. Leung is funded by the President's PhD Scholarships of Imperial College London.}, Jun-Jie Huang and Pier Luigi Dragotti}
\address{Department of Electrical and Electronic Engineering, Imperial College London, UK \\
{\{\href{mailto:chi.leung14@imperial.ac.uk}{chi.leung14}, \href{mailto:j.huang15@imperial.ac.uk}{j.huang15}, \href{mailto:p.dragotti@imperial.ac.uk}{p.dragotti}\}@imperial\emaildot ac\emaildot uk}
}

\begin{document}
%
\maketitle
\begin{abstract}
Finite Rate of Innovation (FRI) theory considers sampling and reconstruction of classes of non-bandlimited continuous signals that have a small number of free parameters, such as a stream of Diracs. The task of reconstructing FRI signals from discrete samples is often transformed into a spectral estimation problem and solved using Prony's method and matrix pencil method which involve estimating signal subspaces. They achieve an optimal performance given by the Cram\'{e}r-Rao bound yet break down at a certain peak signal-to-noise ratio (PSNR). This is probably due to the so-called subspace swap event. In this paper, we aim to alleviate the subspace swap problem and investigate alternative approaches including directly estimating FRI parameters using deep neural networks and utilising deep neural networks as denoisers to reduce the noise in the samples. Simulations show significant improvements on the breakdown PSNR over existing FRI methods, which still outperform learning-based approaches in medium to high PSNR regimes.
\end{abstract}
\vspace*{-0.2cm}
\section{Introduction}
\label{sect:intro}
\vspace*{-0.2cm}
Signals with finite rate of innovation (FRI) have finite degrees of freedom per unit time and include both bandlimited signals as well as non-bandlimited functions. A typical example of FRI signals is a stream of $K$ pulses. It has a $2K$ rate of innovations as the signal can be defined by the amplitudes and the locations of the $K$ pulses. FRI theory \cite{Vetterli2002,Dragotti2007,Blu2008,Urigen2013} has shown that it is possible to perfectly reconstruct classes of non-bandlimited continuous signals with finite rate of innovation from discretised samples. This leads to a wide range of applications, including calcium imaging \cite{Onativia2013}, functional magnetic resonance imaging (fMRI) \cite{Dogan2014} and electrocardiogram (ECG) \cite{Hao2005}.

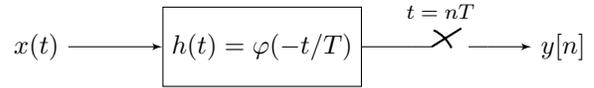
\begin{figure}[!htb]
\centering
\tikzstyle{block} = [draw, rectangle, minimum height=3em, minimum width=6em]
\tikzstyle{input} = [draw=none]
\tikzstyle{output} = [coordinate]

\begin{tikzpicture}[auto, node distance=2cm,>=latex']
    \node [input] at (0,0) (input) {$x(t)$};
    \node [block, right of=input, node distance=3cm] (sampler) {$h(t)=\varphi(-t/T)$};
    \node [coordinate, right of=sampler] (output) {};
    \node [coordinate, right of=output, xshift= -1cm](test){};
    \node [draw=none, right of=test, xshift= -1cm](test1){$y[n]$};

	\draw [->] (input) -- (sampler);
	\draw [-] (sampler) -- (output);
	\draw [-] (output) to[cspst] node[name=t, xshift= -0.5cm, yshift= 0.25cm] {\footnotesize
 $t=nT$} (test);
	\draw [->] (test) -- (test1);

\end{tikzpicture}
\caption{Acquisition process that converts continuous time signal $x(t)$ into discrete time samples $y[n]= \left\langle x(t),\varphi\left({t}/{T}-n\right)\right\rangle$.}
\label{fig:acquisition}
\vspace*{-0.2cm}
\end{figure}

A typical acquisition process involves filtering the input signal $x(t)$ with filter $h(t)=\varphi(-t/T)$ and sampling at a regular interval $t=nT$, as illustrated in \cref{fig:acquisition}. Perfect reconstruction of non-bandlimited FRI signals can be achieved by using a class of kernels $\varphi(t)$ including polynomial and exponential reproducing functions and functions satisfying generalised Strang-Fix conditions \cite{Strang2011,Urigen2013}. The reconstruction task can therefore be transformed into a spectral estimation problem that can be solved by estimation methods such as Prony's method with Cadzow denoising \cite{Prony1795,Cadzow1988} and matrix pencil method \cite{Hua1990} which involve the use of Singular Value Decomposition (SVD) to estimate signal subspaces. Under noisy conditions, it has been found that the reconstruction performance follows the Cram\'{e}r-Rao bound in low noise regime \cite{Cramer1946,Rao1945}. However, the performance breaks down when peak signal-to-noise ratio (PSNR) drops below a certain threshold. It is conjectured to be due to the subspace swap event \cite{Wei2015a} which refers to the confusion of the orthogonal subspace with the signal subspace under noisy conditions.

To avoid this inherent breakdown in subspace-based methods, we propose to solve the FRI reconstruction problem using deep neural networks to learn from training data pairs. We investigate two approaches: (1) using deep neural networks to directly estimate FRI parameters, (2) using deep neural networks to denoise the discrete samples. While there are works utilising deep neural networks to perform spectral estimation on problems such as estimating the frequencies of multisinusoidal signals \cite{Izacard2019,Mathew1994}, estimating the direction of arrival of multiple sound sources \cite{Adavanne2018,Xiao2015}, in this paper, we focus on solving the original reconstruction problem, which is to accurately estimate FRI parameters from discrete samples directly using deep neural networks to avoid the subspace swap event caused by traditional subspace-based methods. 

The rest of the paper is organised as follows: In \Cref{sect:streamofdiracs}, we discuss the occurrence of the breakdown event by illustrating an example of using subspace-based methods to recover a stream of Diracs. We then propose two deep neural network-based approaches to solve the reconstruction problem in \Cref{sect:proposedmethods}. In \Cref{sect:simresults}, we present the simulation results and compare them against the traditional subspace methods. We then conclude in \Cref{sect:conclusion}.

\section{Breakdown in subspace-based methods}
\label{sect:streamofdiracs}
\begin{figure}[!tb]
\centering
\begin{tikzpicture}
    \node[anchor=south west,inner sep=0] (image) at (0,0) {\includegraphics[width=\linewidth]{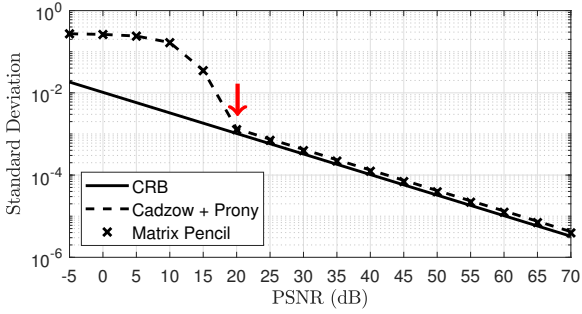}};
    \begin{scope}[x={(image.south east)},y={(image.north west)}]
        \draw [red,ultra thick,->] (0.39,0.7) -- (0.39,0.6);
    \end{scope}
\end{tikzpicture}
\caption{Average standard deviation of retrieved locations of a stream of Diracs over 1000 realisations as compared to the Cram\'{e}r-Rao bound. Both Prony's method with Cadzow denoising and matrix pencil method breaks down when PSNR drops below a threshold (indicated by the red arrow).}
\label{fig:breakdown}
\end{figure}
\begin{figure}[t]
    \centering
    \includegraphics[width=\linewidth]{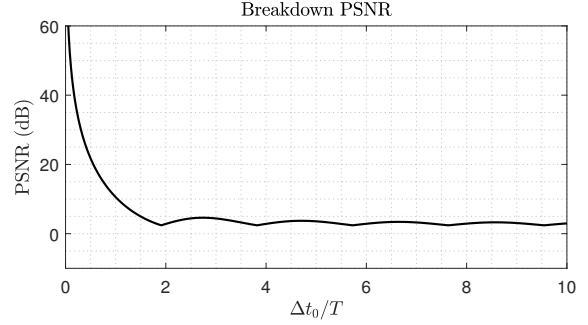}
    \caption{Relationship between breakdown PSNR and the distance between Diracs in the case of $K = 2, N = 21 = P+1\text{ and } \lambda = \frac{2\pi}{N}$ (after \cite{Wei2015a}).}
    \label{fig:breakdownPSNR}
\end{figure}

In this paper, we consider the most basic FRI signal: a stream of $K$ Diracs. We are particularly interested in the performance related to retrieving locations of the $K$ Diracs. To compare our proposed methods with the subspace-based methods in their optimal settings, we consider the reconstruction of a $\tau$-periodic stream of $K$ Diracs: 
\begin{align}
  x(t) = \sum_{l\in\mathbb{Z}}\sum_{k=0}^{K-1} a_k \delta(t-t_k-l\tau)\text{,}
\end{align}
where $\{a_k\in \mathbb{R}\}_{k=0}^{K-1},\{t_k\in \mathbb{R}\}_{k=0}^{K-1}$ are the amplitudes and locations of the Diracs. To sample the continuous signal $x(t)$, we use an exponential reproducing kernel $\varphi(t)$ that can reproduce complex exponentials:
\begin{align}
    \sum_{n\in\mathbb{Z}}c_{m,n}\varphi(t-n) = e^{j\omega_m t}\text{,} 
\end{align}
with $\omega_m = \omega_0 +m\lambda$ for $m = 0,1,...,P$. Assuming sampling period $T = \tau/N$, it is possible to map the obtained samples $y[n]$ into a sum of exponentials: 
\begin{align}
    s[m] & = \sum_{n=0}^{N-1}c_{m,n}y[n] = \sum_{k=0}^{K-1}a_k \sum_{n\in\mathbb{Z}}c_{m,n}\varphi\left(\frac{t_k}{T}-n\right) \nonumber \\
    & = \sum_{k=0}^{K-1}\underbrace{a_k e^{j\omega_0t_k/T}}_{b_k} \left(\underbrace{e^{j\lambda t_k/T}}_{u_k}\right)^m = \sum_{k=0}^{K-1} b_k u_k^m\text{.} \label{eq:nonlinear}
\end{align}
The amplitudes of Diracs $\{a_k\}_{k=0}^{K-1}$ has been mapped to the amplitudes of the exponentials $\{b_k\}_{k=0}^{K-1}$ while the locations of Diracs $\{t_k \}_{k=0}^{K-1}$ have been transformed to $\{u_k \}_{k=0}^{K-1}$. This forms a spectral estimation problem that can be solved using the aforementioned subspace-based methods. Particularly, we are interested in retrieving the locations of Diracs $\{t_k\}_{k=0}^{K-1}$ due to its non-linear nature in the problem seen in \cref{eq:nonlinear}. Previous works \cite{Vetterli2002,Dragotti2007,Blu2008,Urigen2013} have shown that the subspace-based methods achieve an optimal reconstruction performance defined by the Cram\'{e}r-Rao bound until peak signal-to-noise ratio (PSNR) drops below a threshold as shown in \cref{fig:breakdown}. In \cite{Wei2015a}, Wei and Dragotti have conjectured that the reason of the breakdown in subspace-based methods is due to the confusion between noise and signal subspaces in performing spectral estimation. A mathematical relationship \cite{Wei2015a} has then been drawn between this breakdown PSNR and the relative distance between two neighbouring Diracs $\Delta t_k/T$ with $\Delta t_k = t_{k+1} - t_{k}$. For instance, when there are two Diracs of same amplitude, the necessary condition for the subspace swap event is when
\begin{align}
    \text{PSNR} < 10\log_{10}\frac{8\left(\frac{P}{2} + 1\right) \ln\left(\frac{P}{2} +1\right)}{\left(\frac{P}{2}+1-\frac{\sin(\frac{\lambda}{2}(\frac{P}{2}+1)\Delta t_0/T)}{\sin(\frac{\lambda}{2}\Delta t_0/T)}\right)^2}\text{.} \label{eq:breakdownPSNR}
\end{align}
\cref{fig:breakdownPSNR} visualises the breakdown condition in \cref{eq:breakdownPSNR}. It shows that the smaller the distance between two nearby Diracs, the higher the breakdown PSNR will be. Thus, the subspace swap event occurring inherently in current FRI methods stops us from recovering FRI signals with a higher resolution under strong noise.

\section{Proposed Methods}
\label{sect:proposedmethods}
In this paper, we propose to utilise deep neural networks which are learnt using training data pairs as a tool to alleviate the subspace swap problem.

\subsection{Direct inference of FRI parameters from deep neural network}
The first proposed approach is to infer the FRI parameters directly from the noisy samples $\{ \tilde{y}[n] \}_{n=0}^{N-1}$ using a deep neural network. The occurrence of the subspace swap event analysed in \cite{Wei2015a} is inherent to subspace-based reconstruction methods. Hence, to avoid subspace swap event and the eventual performance breakdown, we consider as alternative to learn the transformation from the noisy samples $\{ \tilde{y}[n] \}_{n=0}^{N-1}$ to ground-truth FRI parameters $\{t_k \}_{k=0}^{K-1}$ directly using a deep neural network.

Furthermore, contrary to the traditional subspace methods where the information of the sampling kernel is encoded in $\{c_{m,n}\}$, this approach does not require any explicit information about the sampling kernel $\varphi(t)$. Instead, by training the network with large amount of data from the same sampling kernel, the network aims to obtain this information implicitly.
\subsubsection{Proposed Network Architecture}
\begin{figure}[t]
    \centering
    \includegraphics[width=0.8\linewidth]{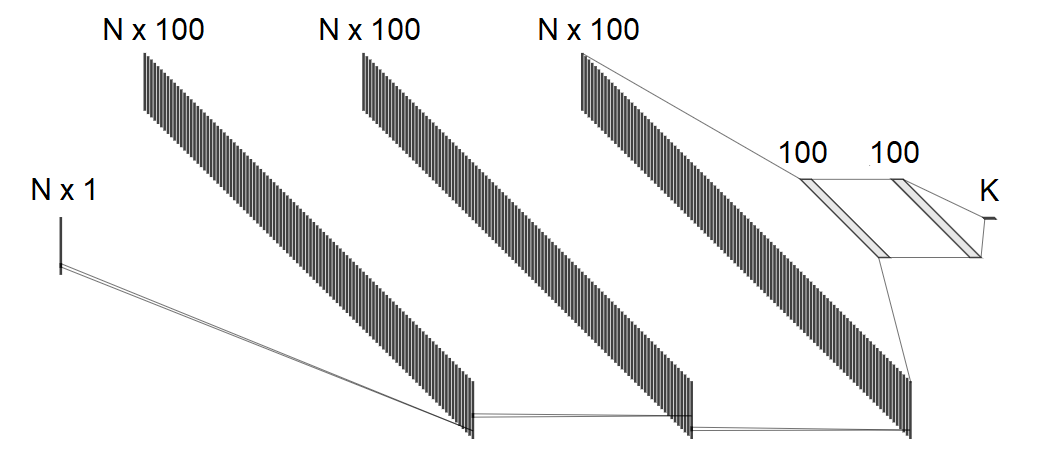}
    \vspace*{-0.25cm}
    \caption{Neural network architecture to perform inference from the observed noisy samples $\{ \tilde{y}[n] \}_{n=0}^{N-1}$ to the locations of Diracs $\{t_k \}_{k=0}^{K-1}$.}
    \label{fig:architecture}
    \vspace*{-0.3cm}
\end{figure}
The neural network consists of 3 convolutional layers followed by 3 fully connected layers as shown in \cref{fig:architecture}. Each of the convolutional layers has 100 filters of size 3. Rectified linear unit (ReLU) is used as the activation function between each two layers. Backpropagation with Adam optimiser \cite{Kingma2014} is used for learning.
\subsubsection{Cost Function}
As the goal of the task is to infer the true locations $\{t_k \}_{k=0}^{K-1}$ from the noisy samples $\{ \tilde{y}[n] \}_{n=0}^{N-1}$, we aim to minimise the discrepancy between the estimated locations $\{\hat{t}_k \}_{k=0}^{K-1}$ and the ground truth locations $\{t_k \}_{k=0}^{K-1}$. Two candidates of the cost function are the absolute differences (L1 loss)
$
f_1 = \sum_{k=0}^{K-1}\left|\hat{t}_k-t_k\right|,
$
and the squared differences (L2 loss)
$
f_2 = \sum_{k=0}^{K-1}\left(\hat{t}_k-t_k\right)^2.
$
Both cost functions yield a similar performance in simulations.

\subsection{Deep neural network as a denoiser}
The second approach is to use deep neural network as a denoiser on the noisy samples $\{ \tilde{y}[n] \}_{n=0}^{N-1}$, followed by a subspace-based method to estimate the FRI parameters. This approach is motivated by the possibility to lower the breakdown PSNR without significantly altering the performance in the low noise regime because the subspace-based methods are still used to perform FRI reconstruction.

The training setting is similar to that used in the direct inference approach. The only modification in the network architecture from \cref{fig:architecture} is the change in size of the 3 fully connected layers to $100N, 20N \text{ and } N,$ respectively. The training loss function is the squared difference between the noisy samples $\{\tilde{y}[n]\}_{n=0}^{N-1}$ and the noiseless samples $\{y[n]\}_{n=0}^{N-1}$.

\section{Simulation Results}
\label{sect:simresults}
\begin{figure*}[!t]
\centering
\begin{subfigure}[b]{0.32\linewidth}
\centering
\hspace*{0.05\linewidth}
\includegraphics[width=\linewidth]{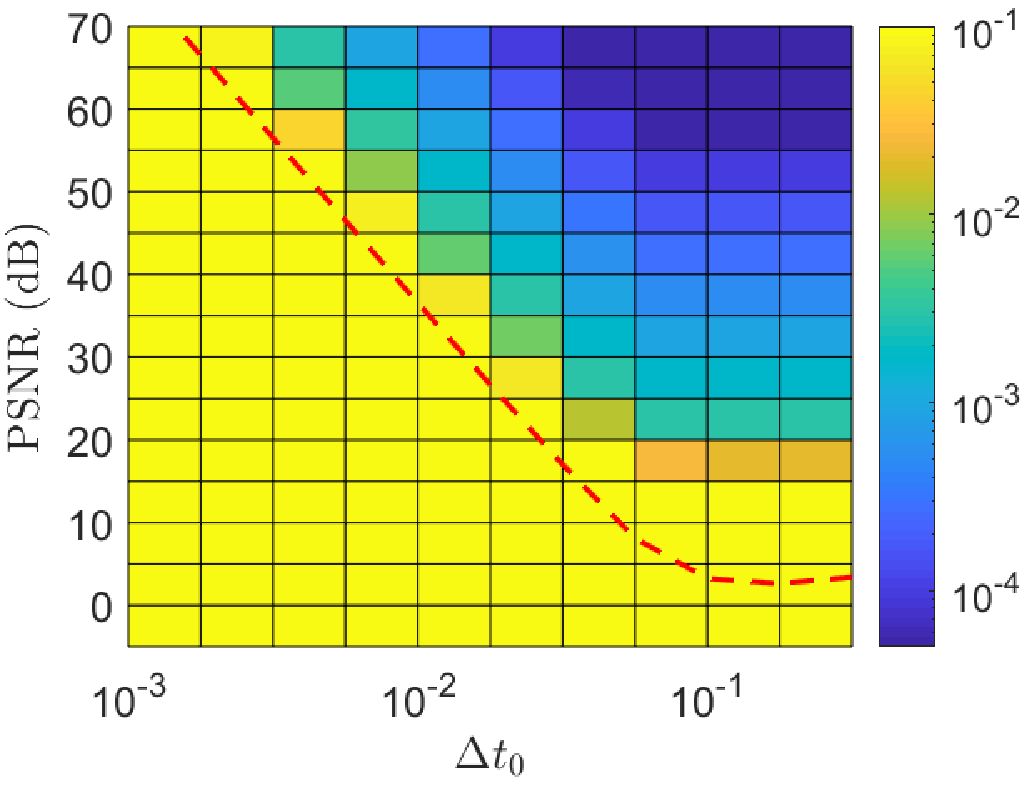}
\caption{Matrix pencil}
\end{subfigure}%
\hfil
\begin{subfigure}[b]{0.32\linewidth}
\centering
\hspace*{0.05\linewidth}
\includegraphics[width=\linewidth]{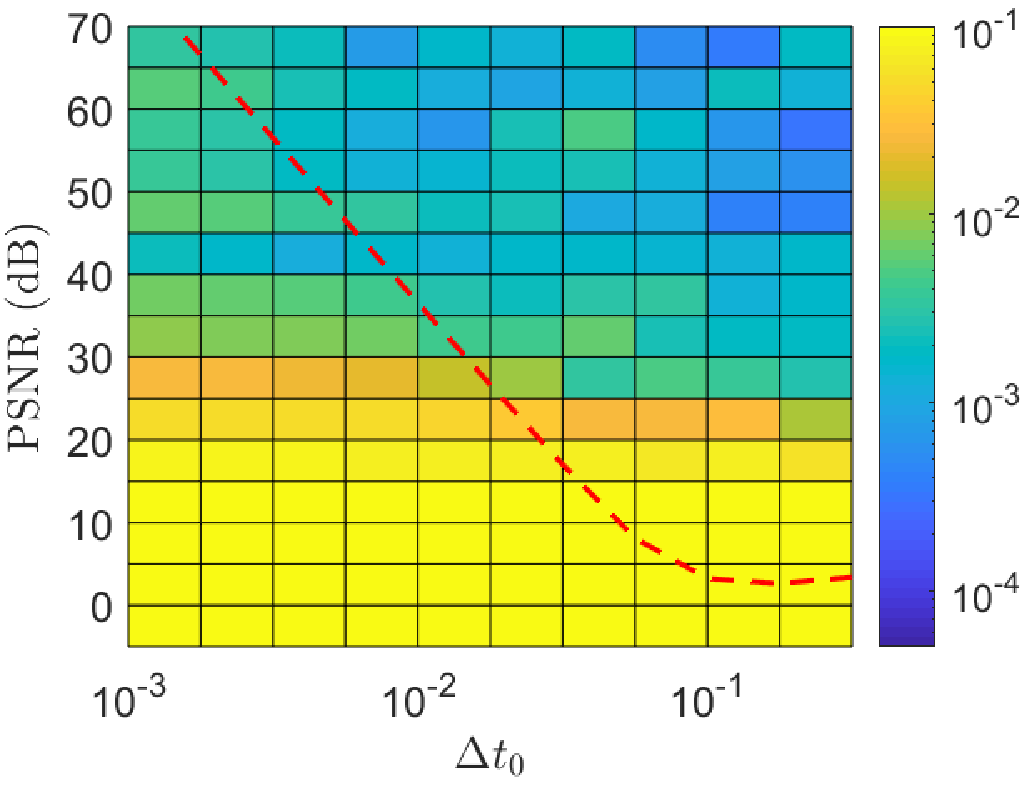}
\caption{Direct inference}
\label{fig:results_direct}
\end{subfigure}
\hfil
\begin{subfigure}[b]{0.32\linewidth}
\centering
\hspace*{0.05\linewidth}
\includegraphics[width=\linewidth]{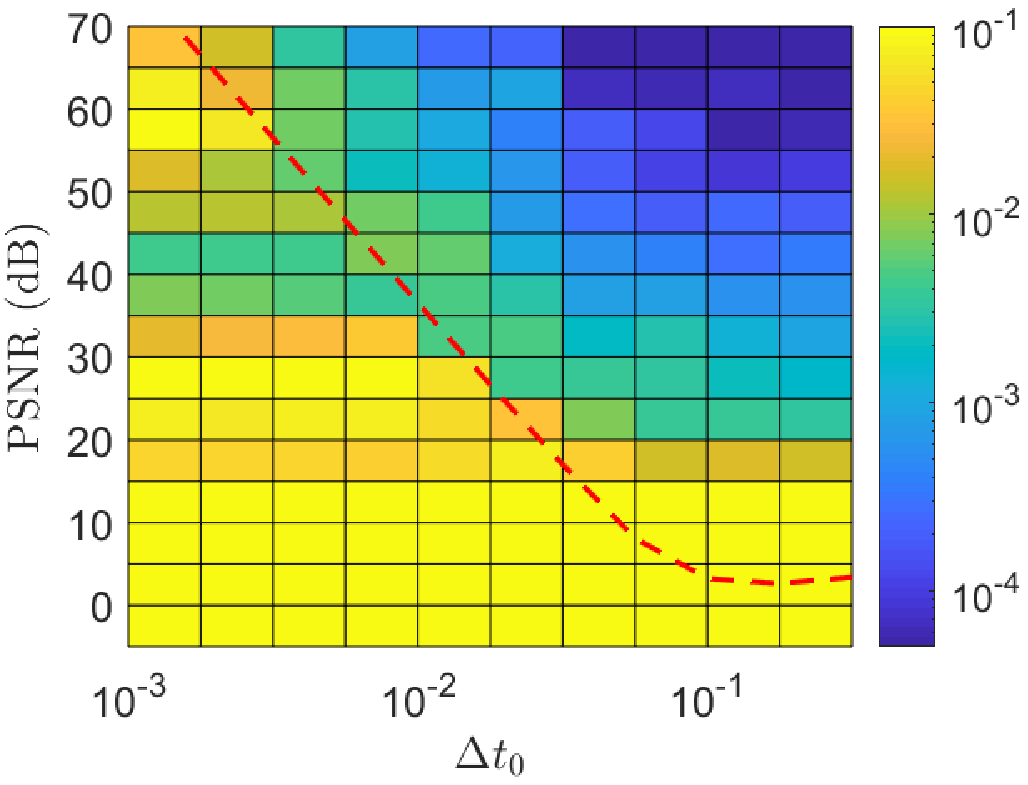}
\caption{Denoiser with matrix pencil}
\end{subfigure}%
\vspace*{-0.3cm}
\caption{Average standard deviation of the retrieved locations of a stream of Diracs ($N=21, K=2$) over 1000 realisations at each PSNR-$\Delta t_0$ pair using different methods. The red dotted line refers to the breakdown PSNR using subspace-based methods shown in \cref{eq:breakdownPSNR} \cite{Wei2015a}.}
\label{fig:results}
\vspace*{-0.4cm}
\end{figure*}
\begin{figure*}[!ht]
\begin{minipage}[t]{\columnwidth}
\centering
\begin{subfigure}[b]{\linewidth}
\centering
\includegraphics[width=\linewidth]{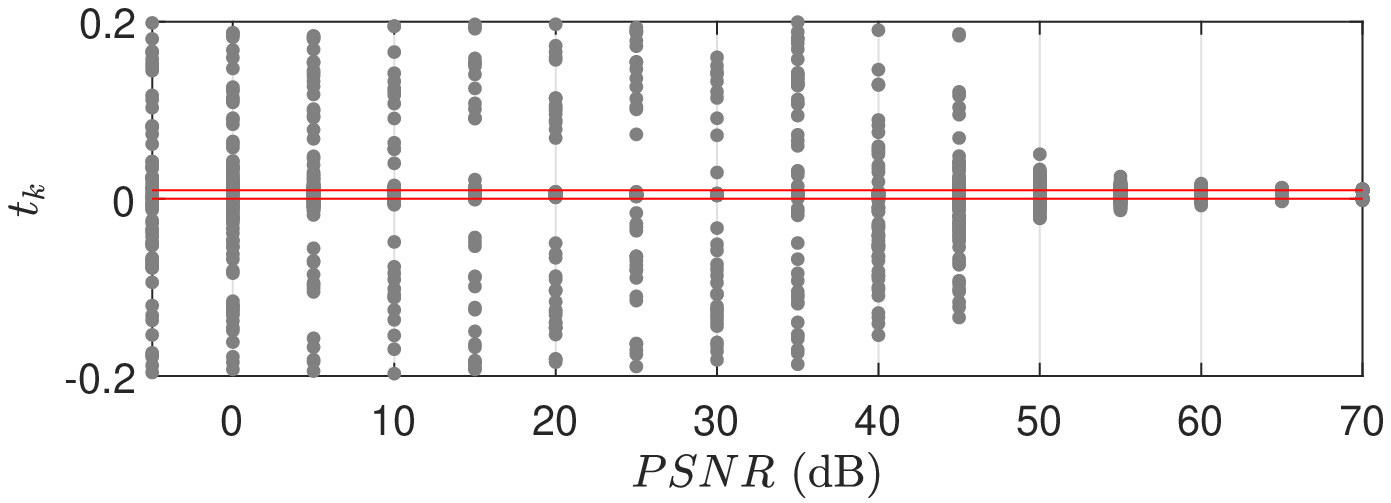}
\caption{Matrix Pencil}
\end{subfigure}%
\\
\begin{subfigure}[b]{\linewidth}
\centering
\includegraphics[width=\linewidth]{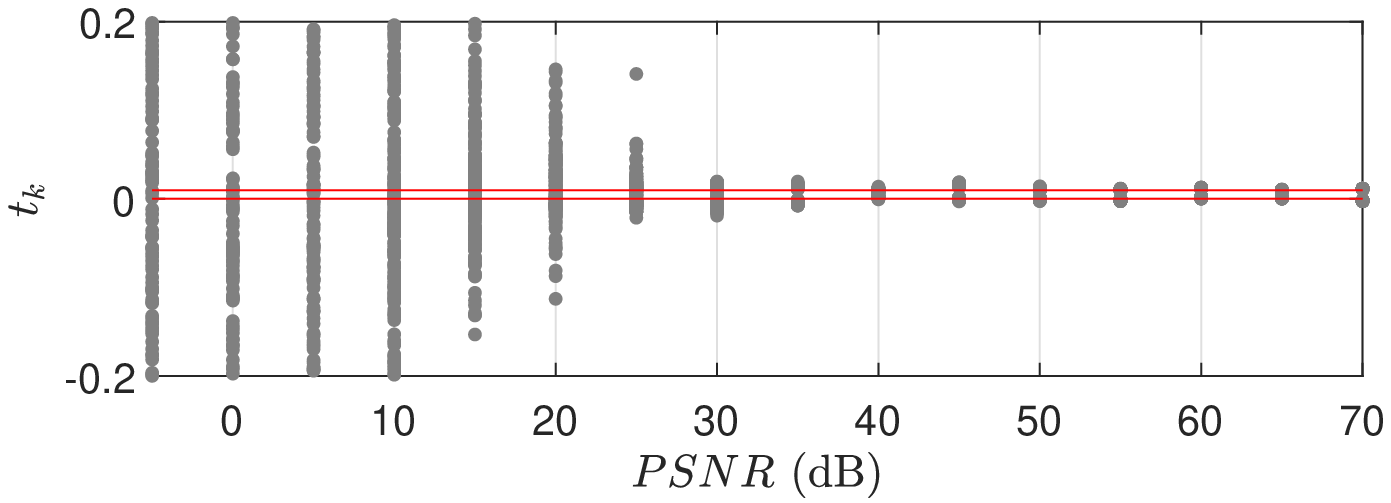}
\caption{Direct inference}
\end{subfigure}
\vspace*{-0.7cm}
\caption{Scatter plot of the retrieved locations over 100 realisations, where the horizontal lines indicate the true locations of the Diracs at $t_0 = 0, t_1 = 10^{-2}$.}
\label{fig:scatter001}
\vspace*{-0.3cm}
\end{minipage}
\hfill
\begin{minipage}[t]{\columnwidth}
\centering
\begin{subfigure}[b]{\linewidth}
\centering
\includegraphics[width=\linewidth]{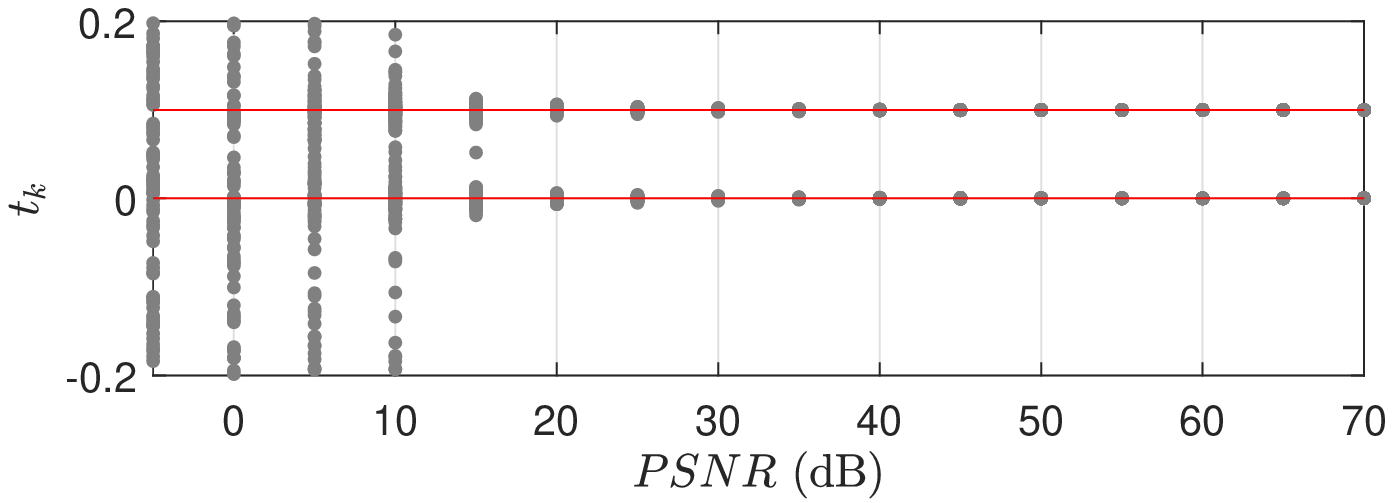}
\caption{Matrix Pencil}
\end{subfigure}%
\\
\begin{subfigure}[b]{\linewidth}
\centering
\includegraphics[width=\linewidth]{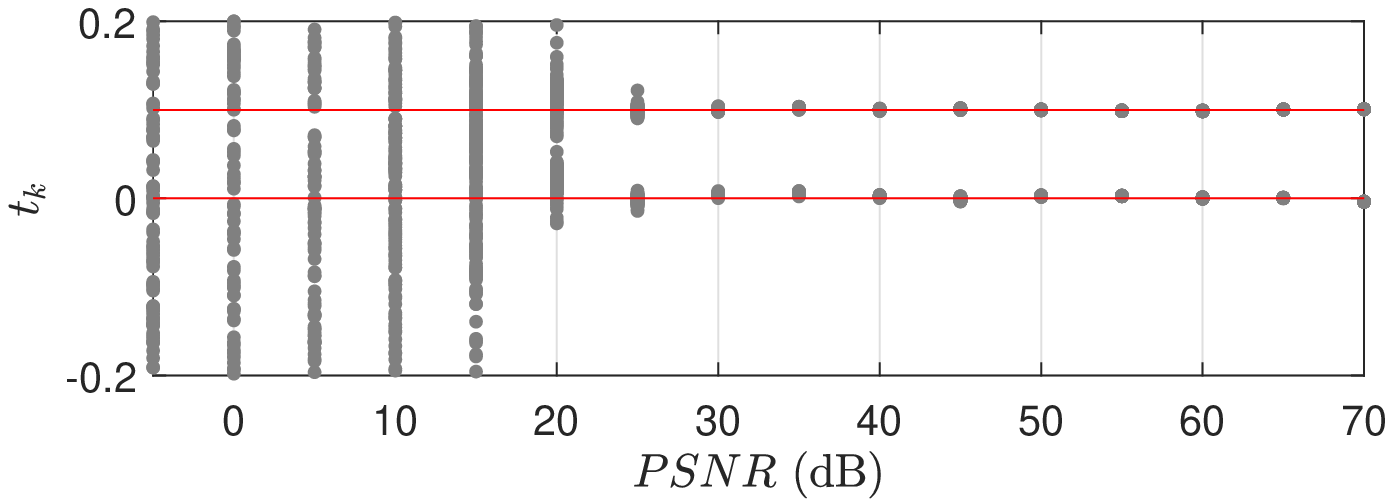}
\caption{Direct Inference}
\label{fig:scatter01_direct}
\end{subfigure}
\vspace*{-0.7cm}
\caption{Scatter plot of the retrieved locations over 100 realisations, where the horizontal lines indicate the true locations of the Diracs at $t_0 = 0, t_1 = 10^{-1}$.}
\label{fig:scatter01}
\vspace*{-0.3cm}
\end{minipage}
\end{figure*}
In this section, we compare the performance of our proposed methods with the subspace-based matrix pencil method \cite{Hua1990}. This is measured by the standard deviation of the retrieved locations of Dirac $t_k$, defined as: 
\begin{align}
f_{sd}=\sqrt{\frac{\sum_{i=0}^{I-1}\left(\hat{t}_k^{(i)}-t_k\right)^2}{I}}\text{,} 
\end{align}
where $\hat{t}_k^{(i)}$ and $I$ are the $i$-th estimation and the number of realisations respectively. For sampling kernel, we consider optimal settings for the subspace-based methods \cite{Wei2015a}, where the sampling kernel $\varphi(t)$ is an exponential reproducing kernel of maximum order and minimum-support (eMOMS) \cite{Urigen2013} that can reproduce $P+1=N$ exponentials at $\omega_0 = \frac{-P\pi}{P+1}$ and $\lambda =\frac{2\pi}{P+1}$. The simulation focuses on a basic setting of reconstructing a stream of 2 Diracs with $t_k \in [-0.5, 0.5)$ and $a_k \in \mathbb{R}^+$. The number of samples and signal period are set to $N=21$ and $\tau = 1,$ respectively. 

To evaluate the reconstruction methods at different levels of noise, the samples $y[n]$ are corrupted with additive white Gaussian noise. A network is trained for every PSNR $\in [-5, 70]$ dB with a step of 5 dB. The training set for each network consist of $10^5$ training data with $t_k \in \mathcal{U}(-0.5,0.5)$ and $a_k \in \mathcal{U}(0.5,10)$ for $k=0,1$ where $\mathcal{U}(a,b)$ denotes uniform distribution between $a$ and $b$.

To investigate the breakdown effect caused by relative distance between two Diracs, we further assume constant amplitudes for two Diracs with $a_0 = a_1 = 2$. We then fix the first Dirac at $t_0=0$ and change $\Delta t_0 \in [10^{-0.5},10^{-3}]$ evenly on a logarithmic scale with a step of $10^{-0.25}$. For each pair of PSNR and $\Delta t_0$, Monte Carlo simulations with 1000 realisations have been applied to evaluate the standard deviation of estimated locations. \cref{fig:results} shows the reconstruction performance of the deep neural network methods and matrix pencil method. Despite the outperformance of the traditional matrix pencil method over the deep neural network approaches when the PSNR is high and the Diracs are sufficiently far apart, we can see that both deep neural network approaches lowers breakdown PSNR, indicated by spread of the low standard deviation region. For instance, when $\Delta t_0=10^{-2}$, both deep neural network-based methods requires PSNR $\geq$ 30 dB whereas matrix pencil method requires PSNR $\geq$ 45 dB. Nonetheless, there exists a discrepancy between the deep neural network methods in the high PSNR breakdown regions, where the deep neural network denoiser fails to push the breakdown PSNR boundary. This could possibly be explained by the low noise power in the samples, which leads to minimal differences between the noisy and ground truth samples. Thus, in spite of outperforming the direct inference approach when the PSNR is high and the Diracs are sufficiently far apart, the denoiser has a limited impact to the breakdown PSNR in the breakdown regions with low noise.

As our goal is to enhance the performance in the breakdown regions of the traditional subspace-based methods, we focus on the distinction between the matrix pencil method and the direct inference method we proposed. We take a closer look by selecting two representative cases when the distance of the Diracs are $\Delta t_0 = 10^{-2}$ and $\Delta t_0 = 10^{-1}$. \cref{fig:scatter001} and \cref{fig:scatter01} shows the respective scatter plot of the estimated locations. We can see that the breakdown PSNR of the matrix pencil method is lower when the Diracs are further apart, while our proposed method has a similar breakdown regardless of the Diracs position. This observation is consistent with the result in \cref{fig:results}. On the other hand, we can also recognise the discrepancy in performance between both methods when the PSNR is high and the Diracs are sufficiently far apart by observing that the centers of the scatters at high PSNR in \cref{fig:scatter01_direct} is not entirely aligned with the true locations. A possible reason could be the generalisation error of the neural network in performing the reconstruction tasks with a high precision. These results suggest that deep neural network is successful in lowering the breakdown PSNR regardless of the locations of the Diracs. Nonetheless, it has a room for improvement when the signal of interest is far from the breakdown PSNR.
\vspace*{-0.3cm}
\section{Conclusion}
\label{sect:conclusion}
\vspace*{-0.2cm}
This paper addresses the breakdown of performance in reconstruction of FRI signals caused by the subspace swap event in traditional subspace-based methods under noisy conditions. We hence proposed two approaches to retrieve the FRI signal by direct inference of FRI parameters and denoising the samples using deep neural networks. Simulation results show that our proposed direct inference method can reconstruct FRI signals at a low PSNR region where the existing FRI methods would break down, yet with a slight performance compromise in high PSNR region.

\vfill\pagebreak
\bibliographystyle{IEEEtran}
\bibliography{Template}

\end{document}